\documentclass[twocolumn,showpacs,preprintnumbers,amsmath,amssymb,floatfix]{revtex4}

\usepackage{graphicx}
\usepackage{dcolumn}
\usepackage{bm}

\usepackage{amsmath}
\usepackage{psfrag}
\usepackage{color}
\usepackage{latexsym,amsfonts}
\usepackage{graphicx}
\providecommand{\href}[2]{#2}   
\definecolor{Blue2}{rgb}{0.,0.,0.8125}
\definecolor{Brown3}{rgb}{0.625,0.25,0.}
\definecolor{Cyan4}{rgb}{0.,0.56,0.56}
\definecolor{Green4}{rgb}{0.,0.56,0.}
\definecolor{LtBlue}{rgb}{0.27,0.42,0.52}
\definecolor{Magenta4}{rgb}{0.5625,0.,0.5625}
\definecolor{Red2}{rgb}{0.8125,0.,0.}

\begin{document}


\title{Exotic Baryons from the Chiral Quark Soliton Model}
\author{Dmitry Borisyuk$^1$}
 \email{borisyuk@ap3.bitp.kiev.ua}
\author{Manfried Faber$^2$}
 \email{faber@kph.tuwien.ac.at}
 
\author{Alexander~Kobushkin$^{1,2}$}%
 \email{akob@ap3.bitp.kiev.ua}
\affiliation{$^1$Bogolyubov Institute for Theoretical Physics, 03143, Kiev, Ukraine}
\affiliation{%
$^2$Atominstitut der \"Osterreichischen Universit\"aten,
        Technische Universit\"at Wien\\
         Wiedner Hauptstr. 8-10, A--1040 Vienna, Austria
}%

\date{\today}

\begin{abstract}
From the interpretation of the  $\Theta^+(1540)$ and $\Xi_{3/2}(1862)$ baryons as an excitation of the ``skyrmion liquid'' with SU(3) flavour symmetry $\overline{10}$ we deduce a new series of baryons, $\Theta_1^{++}$, $\Theta_1^{+}$ and $\Theta_1^0$, situated at the top of the 27-plet of SU(3) flavour, with hypercharge $Y=2$, isospin $I=1$ and spin $J=\frac32$. The $\Theta_1$ mass and width are estimated from the chiral quark soliton model. We demonstrate that the predicted mass, $m_{\Theta_1}=1599~\mathrm{MeV/c^2}$, and broad width are in qualitative conflict with experiment which shows no structure in the total $K^+p$ cross section near $P_\mathrm{lab}=585$~MeV/c. We also study properties of other exotic baryons from the 27- and 35-plets.
\end{abstract}
\maketitle

\section{Introduction\label{sec.1}}
Recently two exotic and narrow baryons, $\Theta^+(1540)$ and $\Xi^{--}_{3/2}(1862)$, which cannot be formed by three quarks were reported. Their simplest quark contents are $uudd\bar s$ and $ddss \bar u$, respectively.  The $\Theta^+(1540)$ baryon was observed in few independent experiments \cite{Nakano,Barmin,Stepanyan,Barth,Asratyan}. Its hypercharge, $Y=2$, follows from strangeness conservation in electromagnetic and strong interactions. No evidence for a $\Theta^+$ partner with electric charge $Q=2$ was found which excludes its isospin $I=1$ \cite{Barth,Kubarovsky}. Despite some arguments that this state may be an isotensor \cite{CPR,Page} we will assume later that the $\Theta^+(1540)$ baryon is an isosinglet.

Due to finite detector resolution experiments \cite{Nakano,Barmin,Stepanyan,Barth,Asratyan} give an upper limit only for the $\Theta^+(1540)$ width, $\Gamma_{\Theta^+} < 9 \div 22$~MeV. Further restrictions (adopting the hypothesis that $\Theta^+(1540)$ is isoscalar) can be obtained from the $K^+d$ total cross section, $\Gamma_{\Theta^+} < 6$~MeV, \cite{Nussinov} and from PWA of $K^+N$ scattering in the $I=0$ channel,  $\Gamma_{\Theta^+} < 1$~MeV, \cite{ASW}. The later estimate agrees with the results of Refs.~\cite{Haidenbauer}~and~\cite{Cahn}.

The $\Xi^{--}_{3/2}(1862)$ baryon with strangeness $S=-2$ was observed in the $\Xi^-\pi^-$ invariant mass spectrum in proton-proton scattering at the CERN SPS \cite{Alt}. The width (including the experimental resolution) was reported with about 18~MeV. The minimal possible isospin of this resonance is $I=\frac32$. In the same experiment another partner of this state, $\Xi^0_{3/2}$, was also observed as a peak in the $\Xi^-\pi^+$ invariant mass spectrum.

Before these experiments were done there had been a lot of theoretical speculations about exotic baryons. Some of the speculations were based on pure quark model calculations \cite{Jaffe,HogaasenS,Strottman,Roiesnel}, others --- on the extended Skyrme soliton model for the SU(3) flavour multiplet $\mu=(0,3)$ with the dimension $N_\mu=\overline{10}$ (anti-decuplet) \cite{Chemtob,BDothan,Praszalowicz,Walliser,DPP,Weigel}.

Three versions of the soliton model calculations \cite{Praszalowicz,DPP,Weigel} predicted the $\Theta^+$ mass, $M_{\Theta^+}^\mathrm{th}=1530~\mathrm{MeV/c^2}$, in excellent agreement with experiment. Moreover, basing on the chiral quark soliton model Diakonov, Petrov and Polyakov \cite{DPP} gave the important qualitative prediction that this state must be narrow, $\Gamma_{\Theta^+}^\mathrm{th} < 15~\mathrm{MeV}$.  Prediction for the $\Xi_{3/2}$ mass  was not so successful ($M_{\Xi_{3/2}}=2070~\mathrm{MeV/c^2}$, \cite{DPP} and 1785~$\mathrm{MeV/c^2}$ in \cite{Praszalowicz03}),  but, in principal, one can reproduce masses of both exotic baryons by an appropriate choice of the model parameters \cite{DP03}. Because these authors assume that both exotic baryons are members of the anti-decuplet the baryon spin-parity is $J^P=\frac12^+$. However it must be mentioned that in principal these states can be also distributed between other penta-quark multiplets (27- and 35-plets) with spin different from $\frac12$. Predictions for higher SU(3) flavour representations were discussed in Refs.~\cite{WKop,BFK}.

Contrary to the picture, where exotic baryons are considered as an excitation of a ``skyrmion liquid''  with appropriate SU(3) flavour symmetry, there were also elaborated approaches based on the constituent quark model \cite{CPR,Page,StankuRiska,Hyogo,JaffeWilczek,KarlinerLipkin,Carlson,Glozman,Bijker,Gerasyuta,DudekClose}. In this pure multi-quark picture $\Theta^+(1540)$ and $\Xi_{3/2}(1862)$ may have quantum numbers different from that predicted by the Skyrme model. The properties of the exotic baryons were also studied in the framework of QCD sum rules \cite{Zhu} and lattice QCD calculations \cite{fodor}.

The aim of the present paper is to study properties (mass spectrum and widths) of penta-quark exotic states from the 27- and 35-plets in the framework of the chiral quark soliton model. We demonstrate that the present experimental information about the exotic baryons gives very strict restrictions on the parameters of the chiral quark soliton model. Using this restrictions we make appropriate predictions for the 27- and the 35-plets. We find that the (chiral quark soliton) model which works perfectly for the octet, decuplet and anti-decuplet fails for higher-dimensional SU(3) flavour representations.

\section{Exotic baryons in the chiral quark soliton model}
In our calculations we use the following Hamiltonian
\begin{equation}\label{Hamiltonian}
\hat H = \hat H_0 + \Delta \hat{H},
\end{equation}
where $\hat H_0$ is the SU(3)-symmetric part and $\Delta \hat{H}$ is responsible for the splitting within SU(3) multiplets.

Using a hedgehog ansatz and assuming rigid rotation in SU(3) space \cite{Witten,Guadagnini} one obtains from the Skyrme Lagrangean the following Hamiltonian for the baryon representation $\mu=(p,q)$ of the SU(3) flavour group
\begin{equation}\label{Hamiltonian0}
\begin{split}
H_0=&M_0 + \frac1{6 I_2}[p^2 + q^2 + pq + 3(p+q)] +\\
& +\left(\frac1{2I_1} - \frac1{2I_2}\right)J(J+1) - \frac{(N_cB)^2}{24I_2},
\end{split}
\end{equation}
where $J$ is the soliton spin, $M_0$ is the energy of the static soliton solution, $I_1$ and $I_2$ are the two moments of inertia, $N_c=3$ is the number of colors and $B=1$ is the baryon number. All quantities $M_0$, $I_1$ and $I_2$ are functionals of the soliton profile.

Some authors (see, e.g., \cite{Cohen}) have expressed doubt about validity of rigid rotation aproximation for penta-quarks. Nevertheless it was shown that such a doubt is ungrounded (see Sect.~3 of Ref.~\cite{DP_disc}).

In the chiral quark soliton model the Hamiltonian $\Delta \hat H$ is chosen phenomenologically \cite{DPP}
\begin{equation}\label{delta_H}
\begin{split}
\Delta \hat{H} = \alpha D^{(8)}_{88}(R) + \beta Y + \frac\gamma{\sqrt3}\sum_{A=1}^3D^{(8)}_{8A}(R)\hat J_A.
\end{split}
\end{equation}
Here $R\in \mathrm{SU(3)}$, $D^{(8)}_{mn}(R)=\frac12 \mathrm{Tr}\left(R^\dag \lambda_m R \lambda_n\right)$ are Wigner rotation matrices for the adjoint SU(3) representation and $Y$ is the hypercharge operator. The constants $\alpha$, $\beta$ and $\gamma$ are related to the current quark masses, $m_u$, $m_d$, $m_s$, the nucleon sigma term and four soliton moments of inertia.

\begin{figure}
  \psfrag{T3}{$I_3$}
  \psfrag{Y}{$Y$}
  \psfrag{Z1}{$\Theta_1$}
  \psfrag{D}{$\Delta,\ N$}
  \psfrag{S}{$\Sigma_2,\ \Sigma,\Lambda$}
  \psfrag{X}{$\Xi_{3/2},\ \Xi$}
  \psfrag{O}{$\Omega_1$}
  \centering
  \includegraphics[height=0.3\textheight]{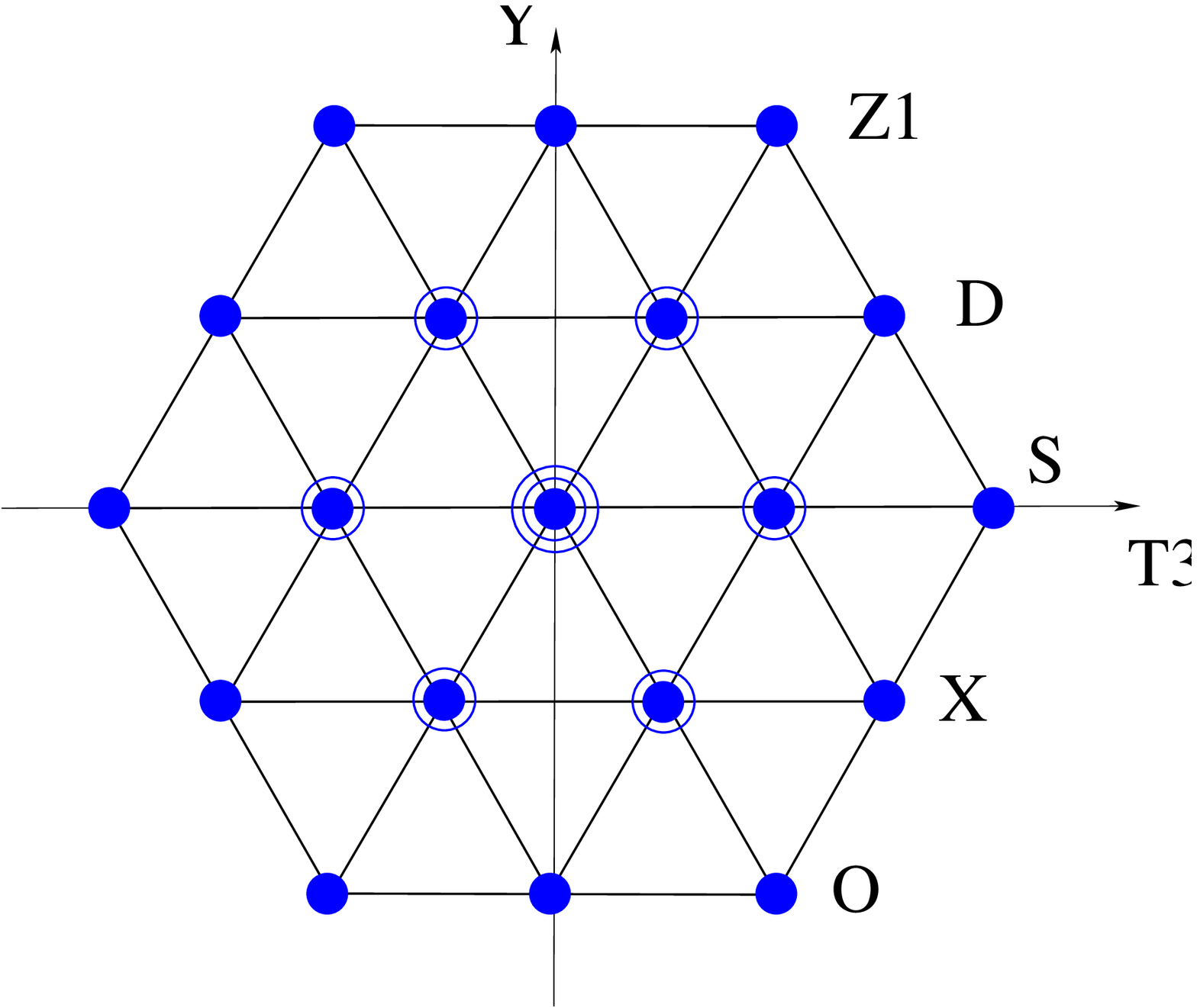}
  \psfrag{A}{$\Theta_2$}
  \psfrag{B}{$\Delta_{5/2},\ \Delta$}
  \psfrag{C}{$\Sigma_2,\ \Sigma$}
  \psfrag{D}{$\Xi_{3/2},\ \Xi$}
  \psfrag{E}{$\Omega_1,\ \Omega$}
  \psfrag{F}{$\Phi$}
  \centering
  \includegraphics[height=0.3\textheight]{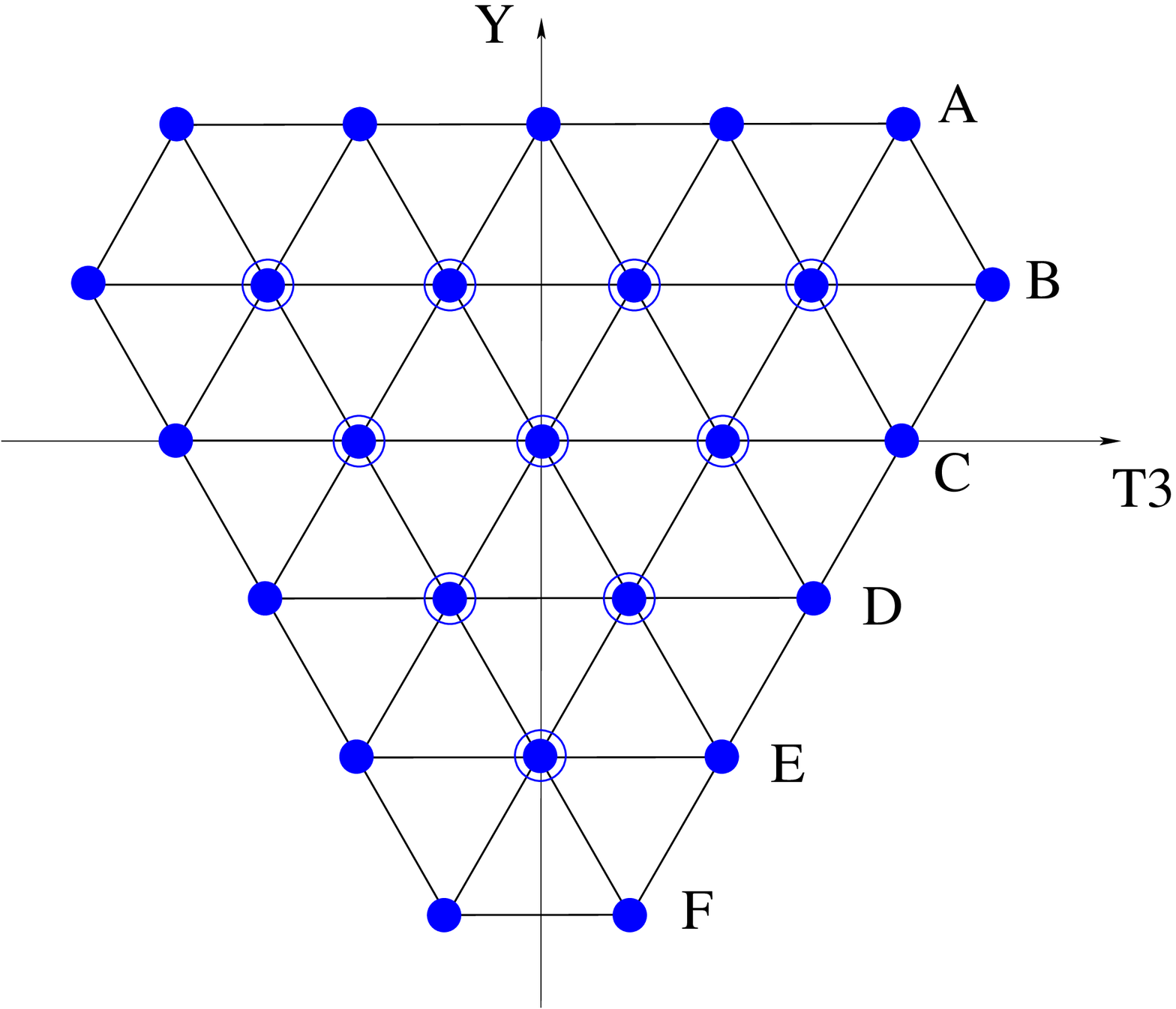}
  \caption{Structure of the 27- (upper figure) and the 35-plet (bottom figure) of baryons in the $I_3Y$ diagram. The number of circles shows the multiplicity.}
 \label{27-plet}
\end{figure}

Due to the Wess-Zumino term the quantization rule selects only such soliton spins $J$, which coincide with one of the allowed isospins $I$ for hypercharge $Y=1$ in the given SU(3) flavour multiplet \cite{Witten,Guadagnini}.
So the lightest irreducible SU(3) representations which can be associated with 3-quark and 4-quark--antiquark systems and appropriate spins $J$ are
\begin{equation}\label{SU(3)-rep}
\begin{split}
\begin{array}{lll}
\text{octet}&\mu=(1,1)&J=1/2\\
\text{decuplet}&\mu=(3,0)&J=3/2\\
\text{anti-decuplet}&\mu=(0,3)&J=1/2\\
\text{27-plet}&\mu=(2,2)&J=1/2 \ \text{or} \ 3/2\\
\text{35-plet}&\mu=(4,1)&J=3/2 \ \text{or} \ 5/2
\end{array}
\end{split}
\end{equation}
The  $I_3 Y$ diagram for the 27- and the 35-plet of baryons is displayed in Fig.~\ref{27-plet}.

The wave functions for baryons with hypercharge $Y$, isospin $I$, isospin 3-projection $I_3$, spin $J$ and its $z$-projection $J_3$ are
\begin{equation}\label{WFunc}
\Psi(R) = \langle R |\mu YII_3JJ_3\rangle = \sqrt{N_\mu}(-1)^{J_3-\frac12}D^{\mu}_{YII_3;1J-J_3}(R),
\end{equation}
where $N_\mu$ is the dimension of the representation $\mu$. Using the first order perturbation theory for the Hamiltonian (\ref{Hamiltonian}) we get the mass spectrum
\begin{equation}\label{sectrum}
\begin{split}
M=&M_0 + \frac1{6 I_2}[p^2 + q^2 + pq + 3(p+q)] +\\
& +\left(\frac1{2I_1} - \frac1{2I_2}\right)J(J+1) - \frac{3}{8I_2} +  \Delta M,
\end{split}
\end{equation}
where
\begin{eqnarray}\label{mass_split}
\Delta M&=&\langle \Delta\hat H \rangle = \\
&=&\langle\mu YII_3JJ_3| \Delta \hat{H} |\mu YII_3JJ_3\rangle. \nonumber
\end{eqnarray}
These splittings are given in Table~\ref{tab:1} together with the results of Ref.~\cite{DPP} for the anti-decuplet.

Exotic states are contained in the $\overline{10}$-, 27- and 35- dimensional representations. For the further analysis we introduce the following universal notations for the exotic states (they are different from the notations used in our previous paper \cite{BFK}):
\begin{itemize}
\item for the states with the hypercharge $Y=2$ we use $\Theta_I$, where the suffix is its isospin; if $I=0$ we use $\Theta$ without suffix;
\item for $Y=1$ we use the notation $\Delta_I$
\item for  $Y=0$ we use $\Sigma_I$
\item for  $Y=-1$ we use $\Xi_I$;
\item for  $Y=-2$ we use $\Omega_I$;
\item for   $Y=-3$ we use $\Phi$.
\end{itemize}

The anti-decuplet contains $\Theta$ and $\Xi_{3/2}$ exotic baryons which due to their $Y$ and/or $I$ values cannot be reduced to three quark systems. From  Fig.~\ref{27-plet} one learns that the 27-plet contains $\Theta _1$, $\Sigma_2$, $\Xi_{3/2}$ and $\Omega_1$ exotic states and the 35-plet contains $\Theta_2$,  $\Delta_{5/2}$, $\Sigma_2$, $\Xi_{3/2}$, $\Omega_1$ and $\Phi$ exotic states. Other members of the $\overline{10}$-, 27- and 35-multiplets are non-exotic with flavour quantum numbers of the octet and decuplet baryons. So these states exist as a mixture of penta-quark and three-quark systems.

\section{Mass spectrum of exotic baryons}

The rotational energy is given by the second and third terms in (\ref{sectrum}). In general it increases very strongly from the octet representation in (\ref{SU(3)-rep}) to the 35 representation. But there is an exception. From numerical results it follows that $I_1>I_2$. This means that in (\ref{sectrum}) the term proportional to $J(J+1)$ becomes more negative for higher angular momenta. So for the 27-plet with $J=\frac32$  and 35-plet with $J=\frac52$ the increase of the rotational energy of the second term in (\ref{sectrum}) can be compensated by the increase of the negative contribution of the third term. For example, estimates with typical parameters for the moments of inertia $I_1$ and $I_2$ show that the rotational energy for the 27-plet with $J=\frac32$ increases by $\approx 100$~MeV only, which, in principal, is of the order of the splitting within the SU(3) multiplet!

\begin{table}
\caption{\label{tab:1}Mass splitting in anti-decuplet with $J=\frac12$, in 27-plets with $J=\frac12$ and $\frac32$
and in 35-plet with $J=\frac52$.}
\begin{ruledtabular}
\begin{tabular}{cccc}
anti-decuplet\footnotemark[1]& $Y$ & $I$ & $\Delta M$\\
\hline
$\Theta$  &  2  & 0   &$ (1/4) \alpha +2\beta -(1/8) \gamma$\\
$N$         &  1  & 1/2 & $ (1/8) \alpha + \beta -(1/16) \gamma$\\
$\Sigma$  &  0  & 1   &                                     0\\
$\Xi_{3/2} $ & -1 & 3/2 &$-(1/8) \alpha - \beta +(1/16) \gamma$\\
\hline
27-plet $J=3/2$\\
\hline
$\Theta_1$  &  2  & 1   & $ (1/7)    \alpha +2\beta -(5/14)   \gamma$\\
$\Delta $& 1& 3/2 & $ (13/112) \alpha + \beta -(65/224) \gamma$\\
$N $        &  1  & 1/2 & $ (1/28)   \alpha + \beta -(5/56)   \gamma$\\
$\Sigma_2$ &   0  & 2   & $ (5/56)   \alpha         -(25/112) \gamma$\\
$\Sigma$ &   0  & 1   &$-(1/56)   \alpha         +(5/112)  \gamma$\\
$\Lambda$ &    0  & 0   &$-(1/14)   \alpha         +(5/28)   \gamma$\\
$\Xi_{3/2}$ & -1  & 3/2 &$-(1/14)   \alpha -  \beta +(5/28)  \gamma$\\
$\Xi$ & -1  & 1/2 &$-(17/112) \alpha -  \beta +(85/224)\gamma$\\
$\Omega_1$  & -2  & 1   &$-(13/56)  \alpha -2\beta +(65/112) \gamma$\\
\hline
27-plet $J=1/2$\\
\hline
$\Theta_1$&  2  & 1   &  $ (17/56)  \alpha +2\beta -(1/112)  \gamma$\\
$\Delta$ & 1& 3/2 &  $ (1/28)   \alpha + \beta -(5/56)   \gamma$\\
$N$  &  1  & 1/2 & $ (137/560)\alpha + \beta +(71/1120)\gamma$\\
$\Sigma_2$&   0  & 2   & $ -(13/56) \alpha         -(19/112) \gamma$\\
$\Sigma$&   0  & 1   & $ (13/280) \alpha         +(19/560) \gamma$\\
$\Lambda$ &    0  & 0   & $ (13/70)  \alpha         +(19/140) \gamma$\\
$\Xi_{3/2}$ & -1  & 3/2 & $-(17/112) \alpha - \beta +(1/224)  \gamma$\\
$\Xi$ & -1  & 1/2 & $ (2/35)   \alpha - \beta +(11/70)  \gamma$\\
$\Omega_1$ & -2  & 1   & $-(1/14)   \alpha -2\beta +(5/28)   \gamma$\\
\hline
35-plet $J=5/2$\\
\hline
$\Theta_2$  &    2 & 2   &$-(1/16)\alpha +2\beta -(77/96)\gamma)$\\
$\Delta_{5/2}$ & 1 & 5/2 &$(11/32)\alpha +\beta -(49/192)\gamma$\\
$\Delta$ & 1 & 3/2 &$-(1/8)\alpha+\beta -(7/16)\gamma$\\
$\Sigma_2$      &0 & 2   &$(3/16)\alpha +(7/96)\gamma$\\
$\Sigma$      &0 & 1   &$-(3/16)\alpha-(7/96)\gamma$\\
$\Xi_{3/2}$     &-1& 3/2 &$(1/32)\alpha -\beta +(77/192)\gamma$\\
$\Xi$     &-1& 1/2 &$-(1/4)\alpha -\beta +(7/24)\gamma$\\
$\Omega_1$      &-2& 1   &$-(1/8)\alpha-2\beta -(35/48)\gamma$\\
$\Omega$      &-2& 0   &$-(5/16)\alpha -2\beta +(21/32)\gamma$\\
$\Phi$          &-3& 1/2 &$-(9/32)\alpha -3\beta +(203/192)\gamma$\\
\end{tabular}
\end{ruledtabular}
\footnotetext[1]{From Ref.~\onlinecite{DPP}.}
\end{table}

In the further analysis the momenta of inertia, $I_1$ and $I_2$, the energy of static skyrmion, $M_0$, and the coefficients $\alpha$, $\beta$ and $\gamma$ are not calculated, but regarded as phenomenological parameters. These parameters were estimated by least square fits to the experimental masses of baryons from octet, decuplet and $\Theta$ and $\Xi_{3/2}$ from $\overline{10}$.  Finally we get the following set of parameters
\begin{equation}\label{Parameters}
\begin{split}
1/I_1&=154.5~\mathrm{MeV}, \quad 1/I_2=403.0~\mathrm{MeV},\\
\alpha&=-602.7~\mathrm{MeV},\
 \beta=-22.3~\mathrm{MeV},\
\gamma=-154.5~\mathrm{MeV},\\
M_0&=790.3~\mathrm{MeV}.
\end{split}
\end{equation}

The estimated masses are given in the left column of Table~\ref{tab:2}. One sees that the model predicts two $\Xi_{3/2}$ states with very small mass difference (less than 20~$\mathrm{MeV}/c^2$) and different spin, $\frac12$ and $\frac32$. So {\it a priori} one cannot exclude the possibility that the observed $\Xi_{3/2}(1862)$ can be a member of the 27-plet with $J=\frac32$. We have also considered all possible billeting of $\Theta$ and $\Xi_{3/2}$ in different multiplets, but their widths can be consistent with experimental restrictions only if they belong to $\overline{10}$.
%
\begin{table}
\caption{\label{tab:2}Mass spectrum and width of exotic baryons.
}
\begin{ruledtabular}
\begin{tabular}{ccll}
Particle          & $J$ &Mass ($\mathrm{MeV/c^2}$) &$\Gamma\ (\mathrm{MeV/c^2})$\\
\hline
\multicolumn{4}{l}{anti-decuplet}\\
\hline
$\Theta$                 & 1/2 & 1540 (input) &  1 (input)\\
$\Xi_{3/2}$              & 1/2 & 1862 (input) & 19   \\
\hline
\multicolumn{4}{l}{27-plet}\\
\hline
$\Theta_1$               & 3/2 & 1599         & 62   \\
$\Sigma_2$               & 3/2 & 1697         & 164  \\
$\Xi_{3/2}$              & 3/2 & 1878         & 137  \\
$\Omega_1$               & 3/2 & 2059         & 128  \\
$\Theta_1$               & 1/2 & 1928         & 271  \\
$\Sigma_2$               & 1/2 & 2271         & 266  \\
$\Xi_{3/2}$              & 1/2 & 2272         & 194  \\
$\Omega_1$               & 1/2 & 2273         & 130  \\
\hline
\multicolumn{4}{l}{35-plet}\\
\hline
$\Theta_2$               & 5/2 & 1839         & 29  \\
$\Delta_{5/2}$           & 5/2 & 1701         & 164  \\
$\Sigma_2$               & 5/2 & 1868         & 127  \\
$\Xi_{3/2}$              & 5/2 & 2035         & 101  \\
$\Omega_1$               & 5/2 & 2203         & 83  \\
$\Phi$                   & 5/2 & 2370         & 71
\end{tabular}
\end{ruledtabular}
\end{table}

\section{Decay widths}

In Sec.~\ref{sec.1} we have already stressed that $\Theta^+(1540)$ are expected to be anomalously narrow. The smallness of its width was shown to be due the to cancellation of coupling constants of different order in $N_c$ \cite{DPP,Praszalowicz230}.

Two-body decay widths of the baryons are calculated by sandwiching baryon-baryon-meson  coupling  between  the in- and out- baryon states. In terms of the collective rotation coordinates $R$ the coupling reads
\begin{equation}\label{BBm-coupling}
\begin{split}
-i\frac3{2m_B}\sum_{A=1}^3 \left[G_0D^{(8)}_{mA}(R) - G_1 \sum_{a,b=4}^8d_{Aab}D^{(8)}_{mb}(R)J_a
\right. \\
 \left. -\sqrt{\frac13}G_2D^{(8)}_{m8}(R)J_A
\right] p_A,
\end{split}
\end{equation}
where $d_{Aab}$ is the SU(3) symmetric tensor, the suffices are $a,b=4,\dots,8$, $A=1,2,3$ and $m=1,\dots,8$ is meson flavour index; $\vec p$ is the meson momentum in the resonance rest frame.

The coupling constants $G_0$ and $G_{1,2}$ have different order in $N_c$
\begin{equation}\label{Nc-expansion}
G_0 \sim N_c^{\frac32}, \qquad G_{1,2} \sim  N_c^{\frac12}.
\end{equation}

In Ref.~\cite{DPP} it was argued that
\begin{equation}\label{G2}
\frac{G_2}{G_0 + \frac12 G_1} \sim 0.01
\end{equation}
and thus estimating baryon widths one can neglect $G_2$. After that the nucleon-pion coupling constant reads \cite{DPP}
\begin{equation}\label{g_pi_NN}
g_{\pi NN}=\frac7{10}\left(G_0 + \frac12 G_1\right)=13.6.
\end{equation}
The ratio $G_1/G_0$ remains unknown and can be restricted only by model-dependent constrains. The authors of Ref.~\cite{DPP}  obtained $\Theta_\Gamma^\mathrm{th}=15$~MeV from the lowest value $G_1/G_0=0.4$ coming from the calculations of \cite{CGPPWW,BPG}. But they have stressed that the $\Theta^+(1540)$ resonance can be much narrower after appropriate choice of the $G_1/G_0$ ratio. In our calculations of the exotic baryon widths we try to bring the $\Theta^+(1540)$ width into accordance with its estimates coming from the analysis of $K^+ N$ and $K^+ d$ scattering data \cite{Nussinov,ASW,Haidenbauer,Cahn}. For the input $\Theta_\Gamma=1$~MeV we obtain $G_1/G_0=1.25$. The results of the calculations are summarized in Table~\ref{tab:2}.

\section{$\Theta_1$ puzzle}

We have already mentioned that the predictions of the chiral quark soliton model, as well as of some quark models, about the lowest excitation of the $\Theta^+(1540)$ baryon are probably in conflict with experimental data for $K^+p$ scattering \cite{Jennings}. Indeed according to its quantum numbers ($I=1,\ J=\frac32$) the contribution of the predicted resonance $\Theta_1(1599))$ in the total $K^+p$ cross-section is
\begin{equation}\label{Tot_cs}
\sigma=\frac\pi{k^2}\frac{(2J+1)}{(2j_1+1)(2j_2+1)}
\frac{B_{K^+p}\Gamma^2_\mathrm{tot}}{\left(E_\mathrm{cm}-M_{\Theta_1}\right)^2+\frac{1}{4}\Gamma_\mathrm{tot}^2},
\end{equation}
where $\Gamma_\mathrm{tot}=\frac{\Gamma_{K^+p}}{B_{K^+p}}$ is the total resonance width and $B_{K^+p}$ is the branching ratio, $k$ is momentum in the c.m. frame and $j_1=\frac12$ and $j_2=0$ are proton and kaon spins, respectively. At the resonance peak it gives
\begin{equation}\label{Tot_cs_peak}
\sigma_{|_\mathrm{peak}}=B_{K^+p}\times\frac{8\pi}{k^2} \approx B_{K^+p} \times 82\ \mathrm{mb}.
\end{equation}
According to its mass and quantum numbers, the $\Theta^{++}_1$ has two decay channels
\begin{equation}\label{channels}
\Theta^{++}_1 \to K^+p \quad \mathrm{and} \quad K \pi N.
\end{equation}
It is difficult to estimate the three-body-decay width, but one may expect that it is strongly suppressed by the phase volume (the mass difference between $\Theta_1$ and the final $K \pi N$ system is only 20~MeV/$\mathrm{c^2}$). Taking as ``mostly optimistic'' branching $B_{K^+p}=0.5$ one obtains 42~mb in the peak against the experimental value of $12 \div 15$~mb for the total $K^+p$ cross section at $P_\mathrm{lab}=585$~MeV/c \cite{BOWEN}.

\section{Conclusions}
In the framework of the chiral quark soliton model we calculated the mass spectrum and two-body widths of all exotic penta-quarks. The model parameters were fixed from the mass splitting in the $\frac12^+$ octet and the $\frac32^+$ decuplet and the assumption that $\Theta^+(1540)$ and $\Xi_{3/2}(1862)$ are members of the anti-decuplet. We show that the model predicts the existence of a new isotriplet of $\Theta$-baryons, $\Theta_1^{++}$, $\Theta_1^+$ and $\Theta_1^0$, with hypercharge $Y=2$ and $J^P=\frac32^+$ and mass $1599~\mathrm{MeV/c^2}$. The triplet of $\Theta_1$ baryons is a member of the 27-dimensional representation of the SU(3) flavour group. The width of this state, in contrast with $\Theta$ from the anti-decuplet, is broad. We demonstrate that this prediction of the chiral quark soliton model is in qualitative conflict with experiment which shows no structure in the total $K^+p$ cross section near $P_\mathrm{lab}=585$~MeV/c.

The authors thank to Andro Kacharava and Eugene Strokovsky for helpful discussions.

\end{document}